\documentclass[twocolumn,superscriptaddress,showpacs,amsmath,amssymb]{revtex4}

\usepackage{graphicx}
\usepackage{dcolumn}
\usepackage{bm}

\begin{document}

\preprint{APS/V. Sih}

\title{Control of Electron Spin Coherence Using Landau Level Quantization in a Two-Dimensional Electron Gas}

\author{V. Sih}
\author{W. H. Lau}
\author{R. C. Myers}
\author{A. C. Gossard}
\affiliation{%
Center for Spintronics and Quantum Information\\
University of California, Santa Barbara, CA 93106}

\author{M. E. Flatt\'e}
\affiliation{%
Department of Physics and Astronomy\\
University of Iowa, Iowa City, IA 52242 }

\author{D. D. Awschalom}
 \email{awsch@physics.ucsb.edu}
\affiliation{%
Center for Spintronics and Quantum Information\\
University of California, Santa Barbara, CA 93106}

\date{\today}

\begin{abstract}
Time-resolved optical measurements of electron spin dynamics in
modulation doped InGaAs quantum wells are used to explore electron
spin coherence times and spin precession frequencies in a regime
where an out of plane magnetic field quantizes the states of a
two-dimensional electron gas into Landau levels. Oscillatory
features in the transverse spin coherence time and effective
g-factor as a function of applied magnetic field exhibit a
correspondence with Shubnikov-de Haas oscillations, illustrating a
coupling between spin and orbital eigenstates. We present a
theoretical model in which inhomogeneous dephasing due to the
population of different Landau levels limits the spin coherence
time and captures the essential experimental results.

\end{abstract}


\maketitle

Electron spins in semiconductors have the potential to form the
basis of emerging spintronics~\cite{wolf01} and quantum
information processing technologies~\cite{awsch02}. While the
dynamics of both the electron spin and its orbital degree of
freedom in a two-dimensional electron gas are well understood,
intricate phenomena may be expected in the presence of spin-orbit
interactions. Here, we present time-resolved optical measurements
of the transverse spin relaxation time $T_{2}^\ast$ and effective
g-factor g* on two-dimensional electron gases (2DEG) in a set of
single InGaAs quantum wells (QW). Both $T_{2}^\ast$ and g* exhibit
oscillations when measured as a function of applied magnetic field
that correspond to features in the magnetoresistance, indicating a
sensitivity to the Landau level filling.

An electron in a magnetic field $B$ has a spin precession
frequency $\Omega_{L}$ = $\text{g*}\mu_{\text{B}}B/\hbar$, where
$\mu_{\text{B}}$ is the Bohr magneton and $\hbar$ is Planck's
constant $h$ divided by 2$\pi$. g* can deviate significantly from
the free electron value g $\sim$ 2.0 due to spin-orbit coupling.
Under the application of a strong out-of-plane magnetic field, the
energy spectrum of a 2DEG becomes quantized into Landau levels, in
which the trajectory of the electrons can be characterized as a
cyclotron orbit with radius $R_{c} = \sqrt{\hbar/eB}$, where $e$
is the charge of an electron. When $B = B_{n} = hn_{2D}/en$, where
$n_{2D}$ is the sheet density and $n$ is an integer indicating the
Landau level index, there are $n$ filled Landau levels. The
spacing between Landau levels is periodic in reciprocal field, and
changing the applied magnetic field changes the filling factor of
occupied Landau levels $\nu = hn_{2D}/eB$.

Previous measurements of electron g-factor in a 2DEG as a function
of Landau level filling have been performed primarily using
electrically-detected electron spin resonance (EDESR), which
records a resonant change in the magnetoresistivity due to an
applied microwave excitation~\cite{stein84, dobers88}. The low
number of electron spins in a 2DEG makes the direct detection of
microwave absorption for conventional ESR
difficult~\cite{nestle97}. Although EDESR studies have yielded a
relation between g* and $n$, the resonance feature was only
observable in a small range of magnetic field where the Fermi
energy is located between spin-split Landau levels~\cite{stein84},
and the line-width measured through the conductivity is not
directly related to the spin coherence time~\cite{dobers88}. The
electron g-factor has also been measured using the coincidence
method~\cite{fang68}, but these transport measurements can be
dominated by exchange interaction~\cite{dobers88}. Here, we
measure the spin dynamics of optically injected electrons using
time-resolved Faraday rotation. This allows us to determine
$T_{2}^\ast$ and g* over a wider range of magnetic fields and
observe oscillations that indicate a dependence on $\nu$.

Electron spin coherence and transport measurements are performed
on a set of single modulation doped In$_{0.2}$Ga$_{0.8}$As/GaAs QW
grown by molecular beam epitaxy. The sample structure is 50 nm
GaAs/30 nm $n$-doped GaAs/20 nm GaAs/7.5 nm
In$_{0.2}$Ga$_{0.8}$As/20 nm GaAs/10 nm $n$-doped GaAs/(001)
semi-insulating GaAs substrate. The doping densities of the
Si-doped layers are $5 \times 10^{16}$ cm$^{-3}$ (sample A); $1
\times 10^{17}$ cm$^{-3}$ (B); $3 \times 10^{17}$ cm$^{-3}$ (C);
$5 \times 10^{17}$ cm$^{-3}$ (D); and $8 \times 10^{17}$ cm$^{-3}$
(E). Since the absorption energies of these quantum wells
(photoluminescence peak at 1.33 eV at temperature $T$ = 5 K) are
lower in energy than the band gap of the GaAs substrate, we can
selectively optically excite and detect electron spin polarization
in the quantum well. Low temperature transport measurements are
performed on samples C, D and E in a magneto-optical cryostat with
magnetic fields up to $B$ = 7 T and reveal clear signatures of
Shubnikov-de Haas oscillations. The samples are patterned with a
standard 4:1 Hall bar geometry and are measured using lock-in
detection with an excitation current of 99 nA at 11 Hz. The
electron sheet densities and mobilities at $T$ = 5 K are $5.4
\times 10^{11}$ cm$^{-2}$ and $3.8 \times 10^{4}$ cm$^{2}$/V s
(C), $6.6 \times 10^{11}$ cm$^{-2}$ and $3.1 \times 10^{4}$
cm$^{2}$/V s (D), and $7.0 \times 10^{11}$ cm$^{-2}$ and $2.4
\times 10^{4}$ cm$^{2}$/V s (E). We determine that the electron
effective mass in sample E is 0.064 $m_{e}$ by fitting the
temperature dependence of the amplitudes of the Shubnikov de Haas
oscillations~\cite{bauer72}. Optical measurements (spot diameter
$\sim$50 $\mu$m) performed on patterned Hall bar structures (mesa
width $\sim$150 $\mu$m) are found to reproduce the results of
unprocessed samples, indicating that the processing has little
effect on the electron spin dynamics of the 2DEG.

\begin{figure}
\includegraphics[width=0.44\textwidth]{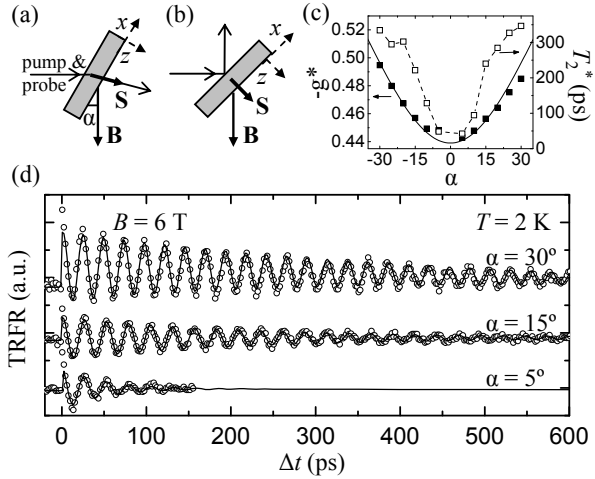}
\caption{\label{fig:epsart} (a) Transmission and (b) 45 degree
reflection measurement geometries. (c) Effective g-factor g*
($\blacksquare$) and spin coherence time $T_{2}^\ast$ ($\square$)
as a function of angle $\alpha$ at $B$ = 6 T and $T$ = 2 K on
sample E. The solid curve is a fit to Eq.~(2). (d) Faraday
rotation as a function of time delay $\Delta$t on sample E at
$\alpha$ = 5, 15 and 30 degrees for $B$ = 6 T and $T$ = 2 K.}
\end{figure}

TRFR, an optical pump-probe spectroscopy, is used to probe the
electron spin dynamics. Using a balanced photodiode bridge and
lock-in detection, rotation angles on the order of 10 microradians
can be measured with sub-picosecond temporal
resolution~\cite{crooker97}. The electron spin magnetization
precesses in the plane perpendicular to the applied magnetic
field, and the Faraday rotation angle as a function of time delay
$\Delta$t can be expressed:
\begin{eqnarray}
\theta_{F}(\Delta t) = A_{1}e^{-\Delta t/T_{1}} + A_{2}e^{-\Delta
t/T_{2}^*} \cos \Omega_{L} \Delta t
\end{eqnarray}
where $A_{1}$ ($A_{2}$) is the amplitude of the electron spin
polarization injected that is parallel (perpendicular) to the
magnetic field, $T_{1}$ is the longitudinal spin coherence time.
Although the sign of g* cannot be determined from such fits,
measurements of In$_{x}$Ga$_{1-x}$As for $0 < x <
0.1$~\cite{weisbuch77} and InAs~\cite{konopka67} indicate that g*
is negative.

Two geometries employed in this measurement are illustrated in
Fig.~1: a transmission geometry in which the [110] direction ($x$)
can be rotated up to $\pm$ 30 degrees from the direction of the
applied magnetic field by an angle $\alpha$ [Fig.~1(a)] and a
reflection geometry where the sample is 45 degrees with respect to
the applied field and the optical paths [Fig.~1(b)]. In the latter
case, the collection path forms a right angle with the incident
light. The sample is mounted so that the magnetic field is in the
($x$, $z$) plane. Figure 1(d) shows TRFR measurements at 6 T and 2
K on sample E. A summary of the angle dependence of $T_{2}^\ast$
and g* is plotted in Fig.~1(c). $T_{2}^\ast$ increases
dramatically with increasing $\alpha$ and out-of-plane magnetic
field; this is related to a suppression of the dominant spin
relaxation mechanism, which is discussed later in the text.

\begin{figure}
\includegraphics[width=0.44\textwidth]{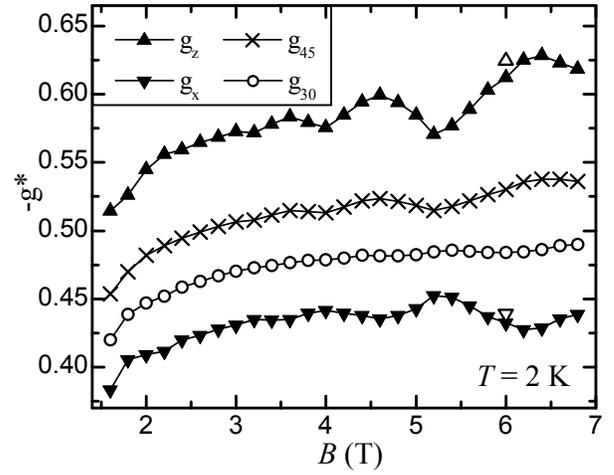}
\caption{\label{fig:epsart} g* measured as a function of magnetic
field $B$ at 45 degrees ($\times$) and 30 degrees ($\circ$) with
respect to the growth direction $z$ for sample E and calculated
values of g$_{z}$ ($\blacktriangle$) and g$_{x}$
($\blacktriangledown$). The hollow symbols show g$_{z}$
($\vartriangle$) and g$_{x}$ ($\triangledown$) as calculated from
the angle dependence fit shown in Fig.~1(c). }
\end{figure}

g* as a function of $\alpha$ can be fit to determine the
components of the g-tensor along the $x$ and $z$
directions~\cite{salisPRB}:
\begin{eqnarray}
|\text{g}_{\alpha}| = \sqrt{\text{g}_{x}^2 \cos^2 \alpha +
\text{g}_{z}^2 \sin^2 \alpha}
\end{eqnarray}
The solid line in Fig.~1(c) is a fit from which $\text{g}_x =
0.663$ and $\text{g}_z = 0.790$. Measurements taken at 30
($\text{g}_{30}$) and 45 degrees ($\text{g}_{45}$) as a function
of field are used to solve for $\text{g}_{x}$ and $\text{g}_{z}$
in Fig.~2. The oscillations in g* are more prominent when measured
in the 45-degree reflection geometry, where a larger component of
the magnetic field is out-of-plane. The results of fitting the
data in Fig.~1(c) to Eq.~(2) are plotted as hollow symbols at $B$
= 6 T for comparison. We account for the discrepancy with an
estimated error in determining $\alpha$ of $\pm$3 degrees. In
order to minimize the effect on $\Omega_{L}$ from the hyperfine
interaction with nuclei~\cite{salisPRL}, a photoelastic modulator
was used to polarize the electron spins, as the time-averaged
electron spin population from a waveplate switching between right
and left circular polarization should be zero. In addition,
measurements are performed at varying lab time intervals in order
to check that the nuclear spins have negligible effect on the
data. Comparisons of $\Omega_{L}$ at positive and negative fields
show the steady-state nuclear field to be less than one percent of
the applied field.

\begin{figure}
\includegraphics[width=0.44\textwidth]{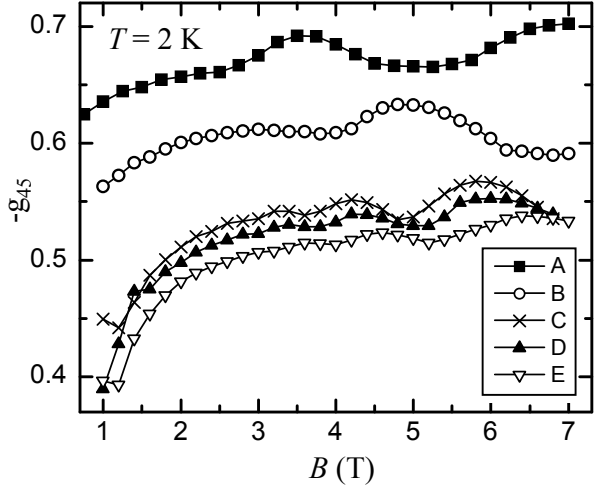}
\caption{\label{fig:epsart} g$_{45}$ measured in the 45 degree
reflection geometry for samples A (barrier doping density $5
\times 10^{16}$ cm$^{-3}$); B ($1 \times 10^{17}$ cm$^{-3}$); C
($3 \times 10^{17}$ cm$^{-3}$); D ($5 \times 10^{17}$ cm$^{-3}$)
and E ($8 \times 10^{17}$ cm$^{-3}$).}
\end{figure}

The field dependence of g* as measured in the 45 degree reflection
geometry ($\text{g}_{45}$) for all five samples is plotted in Fig.
3. All samples exhibit the same qualitative behavior, with the
magnitude of $\text{g}_{45}$ first increasing with magnetic field
and then crossing over to an oscillatory regime at higher field.
As the carrier density is increased from sample A to sample E, the
g-factor increasingly reflects the value of the bulk GaAs g-factor
($-0.44$), indicating enhanced penetration of the electron wave
function into the barriers, while the period of the oscillations
seems to decrease, consistent with the decreasing spacing of the
Landau levels with the increasing sheet density of the 2DEG.

Similarities between g* and Shubnikov-de Haas oscillations in the
45 degree reflection geometry, illustrated with data for sample E
at $T$ = 2 K, 5 K and 20 K in Fig. 4, indicate that g* is
dependent on the filling factor $\nu$. The vertical dotted lines
in Fig.~4 indicate $B_{n}$ for $n =$ 6 to 16. Previous
measurements of spin precession frequencies in a 2DEG using EDESR
established a linear relation between g* and Landau level index
$n$:
\begin{eqnarray}
\text{g}(B,n) = \text{g}_{0} - c(n+\frac{1}{2})B
\end{eqnarray}
where $\text{g}_{0}$ and $c$ are sample dependent constants, but
g* could only be measured in regions of field around odd filling
factors~\cite{dobers88}. Our measurement covers a wider magnetic
field range, revealing oscillatory behavior of g* as a function of
$B$ that tracks the behavior of the Shubnikov-de Haas
oscillations. We fit our data in regions near full filling to
obtain $\text{g}_{0} = 0.405$ and $c = 0.00314$ for our sample and
plot the calculated g-factor dependence in Fig. 4(a) for $n =$ 4
to 12 (dashed lines). The temperature dependence of the TRFR data
demonstrates that the amplitude of the oscillations in g*
diminishes as the temperature is increased from 2 K to 20 K.
Likewise, the Shubnikov-de Haas oscillations, evident in the
magnetic field range presented here at 2 K and 5 K, are faint
below 5 T at 20 K. We observe from power dependences of our
measurement that the data presented here is in a regime where the
number of optically injected carriers does not change the
g-factor, indicating a minimal effect of the pump-probe
measurement on the Fermi level.

\begin{figure}
\includegraphics[width=0.44\textwidth]{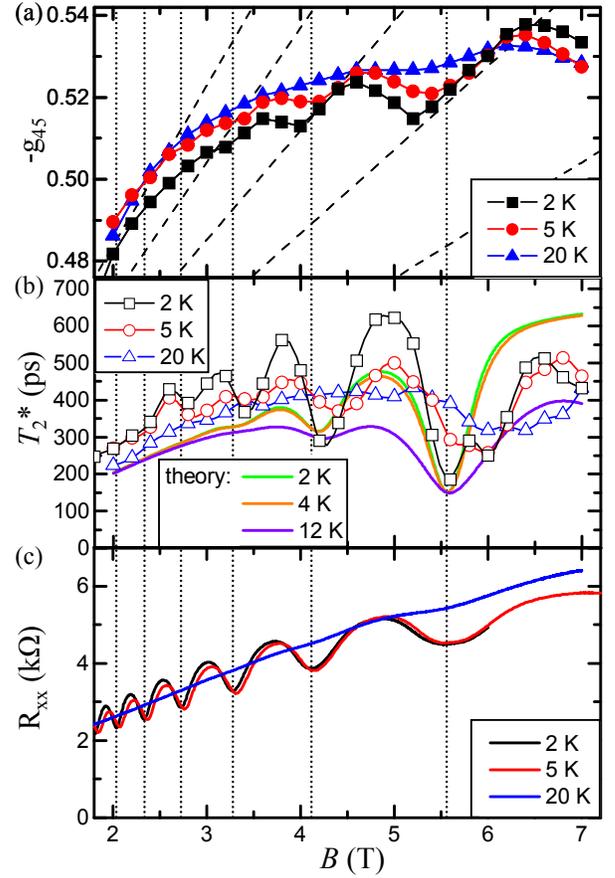}
\caption{\label{fig:epsart} (a) g* measured in the 45 degree
reflection geometry for sample E as a function of $B$ at $T$ = 2
K, 5 K and 20 K (symbols). Also plotted (dashed lines) is
$\text{g}(B, n) = \text{g}_{0} - c(n + \frac{1}{2})B$, with
$\text{g}_{0}$ = 0.405 and $c$ = 0.00314 for $n$ = 4 to 12. (b)
$T_{2}^\ast$ measured (symbols) as a function of B at $T$ = 2 K, 5
K and 20 K and calculated (lines) from a spin relaxation model at
$T$ = 2 K, 4 K and 12 K. (c) $R_{xx}$ as a function of $B$ at $T$
= 2 K, 5 K and 20 K. Also shown are dotted lines indicating the
position of $B_{n}$ for $n$ = 6 to 16.}
\end{figure}

While a dependence in the g-factor on Landau level occupation has
been observed previously~\cite{dobers88}, oscillatory behavior in
$T_{2}$ has not been reported before. $T_{2}^\ast$, as measured in
the 45 degree reflection geometry at 2 K, 5 K and 20 K, is plotted
in Fig.~4(b). From the data, we observe that at low field,
$T_{2}^\ast$ increases quadratically and at high field,
$T_{2}^\ast$ exhibits oscillations in magnetic field whose minima
correspond to $B_{n}$. We next discuss a theoretical model which
explains the dependence of the spin coherence time on magnetic
field. This model calculates the spin relaxation rate $T_{2}^{-1}$
by considering three contributions: a quadratic fit at low field
reflecting the suppression of the D'yakonov-Perel' spin relaxation
mechanism~\cite{dyakonov}, a constant background spin relaxation
rate of 1.2 ns$^{-1}$, and a variance of g-factor mechanism, which
employs the results of a quantitative calculation based on a
generalized ${\bf K \cdot p}$ envelope function theory solved in a
fourteen-band restricted basis set~\cite{wayne}. In the absence of
an applied magnetic field, the D'yakonov-Perel' (DP) spin
relaxation rate $T_{2}^{-1}$ is
\begin{eqnarray}
T_{2}^{-1} = \Omega^2 \tau_{o}
\end{eqnarray}
where $\Omega$ is the precession frequency about the internal DP
field and $\tau_{o}$ is the orbital coherence time. As is
consistent with the DP mechanism, the application of an external
magnetic field increases the spin coherence time by a factor that
is quadratic in applied magnetic field~\cite{meier}
\begin{eqnarray}
T_{2}(B) = T_{2}(0)(1 + a^2 B^2)
\end{eqnarray}
A fit to the data taken at 2 K for the magnetic field range 1 -
2.6 T yields $T_{2}(0)$ = 57 ps and $a$ = 0.96 $\text{T}^{-1}$.
This is the reason for the strong dependence of $T_{2}^\ast$ with
$\alpha$ in Fig.~1(c). The Elliot-Yafet mechanism is less
sensitive to external magnetic field~\cite{EY}. Above 3 T,
$T_{2}^\ast$ exhibits an oscillatory dependence on field that
tracks the Shubnikov-de Haas oscillations. These oscillations are
related to inhomogeneous dephasing of the spin coherence due to
the changing occupation of the Landau levels with magnetic field.
If the width of the Landau levels is comparable to the Landau
level spacing, there will be a number of partially occupied Landau
levels; this occupation will change with field as the spacing
between Landau levels increases. Since electrons in different
Landau levels have different spin precession frequencies, the
$T_{2}^\ast$ that we measure can be dominated by the variance in
the g-factor destroying the phase coherence of the optically
injected spin magnetization. For the variance in g mechanism,
$T_{2}^{-1} \propto \langle\delta\text{g}^2\rangle\tau_{o}$, where
$\langle\delta\text{g}^2\rangle$ is the variance of g. For the
calculations shown in Fig.~4(b), the inhomogeneous broadening of
the Landau levels is 2.6 meV and $\tau_{o}$ = 360 ps. This orbital
coherence time is surprisingly long but may be due to the
importance of localized states located energetically between the
Landau levels. In addition, the calculations appear to
underestimate the oscillation magnitude of g itself.

Another contribution to the oscillatory behavior in $T_{2}^\ast$
may be related to the changing density of states at the Fermi
level, which would lead to a magnetic field dependence of the
scattering time~\cite{burkov}. In our data in Fig.~4(b), however,
the minima of $T_{2}^\ast$ correspond with minima in R$_{xx}$ and
thus the maxima of the conductivity. When the conductivity is
largest, the density of states is largest and the scattering time
is smallest~\cite{ando, ando74}. Thus, from Eq.~(4), $T_{2}^\ast$
should be at a maximum when the resistance is a minimum from the
Ref.~\onlinecite{burkov} model.

The amplitude of the oscillation in the spin coherence time
decreases with increasing temperature. As the temperature
increases, the width of the Landau levels increases, which causes
the features in g* and $T_{2}^\ast$ to become less distinct, both
in the measurements and calculations.

In summary, we have measured the electron spin precession
frequencies and spin coherence times as a function of
perpendicular magnetic field and observed oscillatory features
that indicate a dependence on Landau level quantization.
Measurements were performed on samples of varying doping densities
at a variety of temperatures and magnetic field. The effective
g-factor g* in semiconductors varies widely for materials as it
exhibits a strong dependence on the band gap energy and spin-orbit
coupling; here we have explored the effect of orbital quantization
on g*. The spin coherence time also exhibits an oscillatory
dependence on Landau level filling which may be dominated by
inhomogeneous dephasing. These oscillations are qualitatively
consistent with calculations of $\langle\delta\text{g}^2\rangle$
for this system. The results indicate a possible pathway towards
spin manipulation using orbital quantization; electrical control
of the carrier density could be used to change the Landau level
filling
in a fixed magnetic field with dramatic effects on the g-factor and spin coherence time.\\
\begin{acknowledgments}
We wish to acknowledge the support of DARPA and ARO MURI.
\end{acknowledgments}


\begin{thebibliography}{99}
\bibitem{wolf01} S. A. Wolf, D. D. Awschalom, R. A. Buhrman,
J. M. Daughton, S. von Molnar, M. L. Roukes, A. Y. Chtchelkanova,
and D. M. Treger, Science {\bf 294}, 1488 (2001).
\bibitem{awsch02} {\it Semiconductor Spintronics and Quantum
Computation, NanoScience and Technology}, edited by D. D.
Awschalom, N. Samarth, and D. Loss (Springer-Verlag, New York
2002).
\bibitem{stein84} D. Stein, G. Ebert, K. von Klitzing, and G.
Weimann, Surf. Sci. {\bf 142}, 406 (1984).
\bibitem{dobers88} M. Dobers, K. von Klitzing, and G. Weimann, Phys.
Rev. B {\bf 38}, 5453 (1988).
\bibitem{nestle97} N. Nestle, G. Denninger, M. Vidal, C.
Weinzierl, K. Brunner, K. Eberl, and K. von Klitzing, Phys. Rev. B
{\bf 56}, R4359 (1997).
\bibitem{fang68} F. F. Fang and P. J. Stiles, Phys. Rev. {\bf
174}, 823 (1968).
\bibitem{bauer72} G. Bauer and H. Kahlert, Phys. Rev. B {\bf 5},
566 (1972).
\bibitem{crooker97} S. A. Crooker, J. J. Baumberg, F. Flack, N.
Samarth, and D. D. Awschalom, Phys. Rev. B {\bf 56}, 7574 (1997).
\bibitem{weisbuch77} C. Weisbuch and C. Hermann, Phys. Rev. B {\bf
15}, 816 (1977).
\bibitem{konopka67} J. Konopka, Phys. Lett. A {\bf 26}, 29 (1967).
\bibitem{salisPRB} G. Salis, D. D. Awschalom, Y. Ohno, and H.
Ohno, Phys. Rev. B {\bf 64}, 195304 (2001).
\bibitem{salisPRL} G. Salis, D. T. Fuchs, J. M. Kikkawa, and D. D.
Awschalom, Phys. Rev. Lett {\bf 86}, 2677 (2001).
\bibitem{dyakonov} M. I. D'yakonov and V. I. Perel', Zh. Eksp.
Teor. Fiz {\bf 60}, 1954 (1971) [Sov. Phys. JETP {\bf 33}, 1053
(1971)].
\bibitem{wayne} W. H. Lau, J. T. Olesberg, and M. E. Flatt\'e,
cond-mat/0406201 (2004).
\bibitem{meier} {\it Optical
Orientation, Modern Problems in Condensed Matter Science}, edited
by F. Meier and B. P. Zachachrenya (North-Holland, Amsterdam,
1984), Vol. 8.
\bibitem{EY} R. Elliott, Phys. Rev. {\bf 96}, 266 (1954).
\bibitem{burkov} A. A. Burkov and L. Balents, Phys. Rev. B {\bf
69}, 245312 (2004).
\bibitem{ando} T. Ando, A. B. Fowler, and F. Stern, Rev. Mod.
Phys. {\bf 54}, 437 (1982).
\bibitem{ando74} T. Ando, J. Phys. Soc. Jpn {\bf 37}, 1233 (1974).

\end{thebibliography}
\end{document}